# Observation of superconducting current in a mesa-heterostructure with an interlayer of the strontium iridate film with strong spin-orbit interaction

A.M. Petrzhik[1], K.Y. Constantinian[1], G.A. Ovsyannikov[1], A.V. Shadrin[1,2], A.S. Grishin[1], Yu.V. Kislinski[1], G. Christiani[3], G. Logvenov[3]

[1]Kotel'nikov IRE RAS, Mokhovaya 11-7, Moscow, 125009, Russia

[2] Moscow Institute of Physics and Technology, Dolgoprudny, Moscow Region, 141701, Russia

[3]Max Planck Institute for Solid State Research, Stuttgart, 70569, Germany

The superconducting current has been observed in mesa-heterostructures Nb/Au/Sr$_2$IrO$_4$/YBa$_2$Cu$_3$O$_x$ with Sr$_2$IrO$_4$ interlayer thickness $d$=5 and 7 nm and in-plane sizes $L$=10-50 μm. A strontium iridate, Sr$_2$IrO$_4$, is known as a canted antiferromagnetic insulator at low temperatures and characterized also by the strong spin-orbit interaction due to the impact of the IrO$_2$ plane. The superconducting critical current density $j_C \approx 0.3$ A/cm$^2$ for the case $d$=7 nm was observed at $T$=4.2K. The temperature dependences of the superconducting critical current $I_C(T)$ and the voltage position on the I-V curve of the gap singularity of the Nb electrode $V_\Delta(T)$ show an increase with decreasing temperature and corresponds to the expected BCS behavior of the Nb energy gap $\Delta_{Nb}(T)$. The critical current is very sensitive to the influence of an external magnetic field and reduces twice at an external magnetic field ($H\approx 0.2$ Oe for $L$=40-50 μm) comparable with the earth magnetic field; The magnetic field dependence $I_C(H)$ at low $H$ was narrower than the Fraunhofer pattern about 1.5 times. Both the integer and fractional Shapiro steps at voltages $V_{m,n}$=(m/n)(h/2e)$f_e$ were observed under microwave radiation at frequencies $f_e$=38 GHz and $f_e$=50 GHz. Fractional Shapiro steps (m/n=1/2, 3/2) may point on the presence of the second harmonic in the superconducting current-phase relation.

The spin-triplet superconducting pairing induced at the interface of a superconductor and a material with the strong spin-orbit interaction (SOI) gives promising opportunities for superconducting spintronics applications [1-3]. It has been shown theoretically, that the SOI in a ferromagnet of the superconductor-ferromagnet-superconductor junction can result in a spin-triplet pairing [4-7]. Rashba proximity states in superconducting tunnel junctions, tuning of a magnetization, the proximity-induced triplet superconductivity, and the anomalous Josephson effect have been predicted as well [8-17]. Coulomb repulsion and Pauli quenching affect the superconducting current in tunnel junction with SOI interlayer but do not destroy the possibility to tune it by changing the strength of the spin-orbit interaction [18-21]. So, a lot of very interesting and important for superconducting transport features were predicted theoretically for superconducting structures with an interlayer having strong spin-orbit interaction.

An attempt of the experimental observation of the long-range proximity effect induced by SOI was reported for structures with high-Z (metallic Pt) inclusion into the ferromagnetic interlayer [22]. The unconventional proximity effect was observed in the bridge type Nb junction with a magnetically doped topological insulator (Fe-Bi$_2$Te$_2$Se) and splitting of the zero bias conducting peak and the microwave affected oscillation have been reported [23]. Superconductor-semiconductor, InAs, hybrids were studied in [24] in which the two-dimensional electron gas (2DEG) appears due to the strong SOI. The appearance of the 4$\pi$-Josephson effect which is caused by Majorana surface states in time-reversal-invariant Weyl and Dirac semimetals [25, 26], or in a system with a crossover of the 3D topological insulator to the 2D limit [27] were theoretically suggested. An existence of the 4$\pi$-periodic contribution in the superconducting current phase relation can be responsible for the disappearance of odd steps [11, 13]. Recent investigations of the interplay of a superconductivity, SOI, and Zeeman splitting in the two-dimensional topological insulator HgTe have shown promising results [28] stimulating further studies of Josephson effects in the superconducting structure with SOI interlayer comprising from a more conventional material. The most experimental investigations of the superconducting structure with SOI interlayer were performed in bridge type structure with semiconductor (topological insulator) interlayer. The sandwich type structure is preferable due to the possibility to realize a few nm long distance between superconductors for the interference of superconducting wave function in Josephson junction with a barrier interlayer with SOI.

A natural choice for the interlayer material with a strong spin-orbit interaction in sandwich type junction is the 5d transition metal oxide Sr$_2$IrO$_4$ from the layered Ruddlesden-Popper series (Sr$_{i+1}$Ir$_i$O$_{3i+1}$; i = 1, 2, and $\infty$), in which, along with the electron-electron interaction, a strong SOI is also observed [29-32]. The compound Sr$_2$IrO$_4$ (i =1) is a canted antiferromagnetic insulator with the band splitting and J$_{eff}$ = 1/2 [33]. The intrinsic crystal field splits the degenerate states of 5d electrons into e$_g$ and t$_{2g}$ bands, and the partially filled t$_{2g}$ band splits into J$_{eff}$ = 3/2 and J$_{eff}$ = 1/2 due to the strong SOI over the iridium ions [34]. Unconventional properties of Sr$_2$IrO$_4$, and the realization of interfaces with other oxides, particularly with a superconducting cuprate, are discussed in [35-37]. Moreover, very rich physics of the canted antiferromagnetic insulator Sr$_2$IrO$_4$ [38-40] makes it attractive for the realization of a spin manipulation in the junction with the SOI barrier. However, there is a lack of knowledge of properties, including high-frequency dynamics as the ac Josephson effects, in sandwich type junctions with an oxide barrier with strong SOI.

This paper presents experimental results of superconducting and electron transport characteristics of hybrid micrometer size superconducting Nb/Au/Sr$_2$IrO$_4$/YBa$_2$Cu$_3$O$_x$ mesa-heterostructures with nanometer thickness of the interlayer having strong spin-orbit interaction.

The thin bilayer of YBa$_2$Cu$_3$O$_x$ (YBCO) and Sr$_2$IrO$_4$ (SIO) were grown epitaxially by pulsed-laser deposition (PLD) on (110) NdGaO$_3$ (NGO) single-crystalline substrates. A KrF-excimer laser with frequency 10Hz and 1.6J/cm$^2$ energy density was used to ablate materials from YBa$_2$Cu$_3$O$_x$ and Sr$_2$IrO$_4$ stoichiometric targets. First YBCO thin film was deposited at 830 °C in an oxygen atmosphere of 0.5 mbar. The SIO thin film of required thickness was deposited at 700° C in an argon atmosphere with pressure 0.5 mbar. The substrate temperature was controlled by the radiation pyrometer. The obtained heterostructures were cooled to 500° C and annealed for 30 min in 1 bar of an oxygen atmosphere, before the final cooling to room temperature.

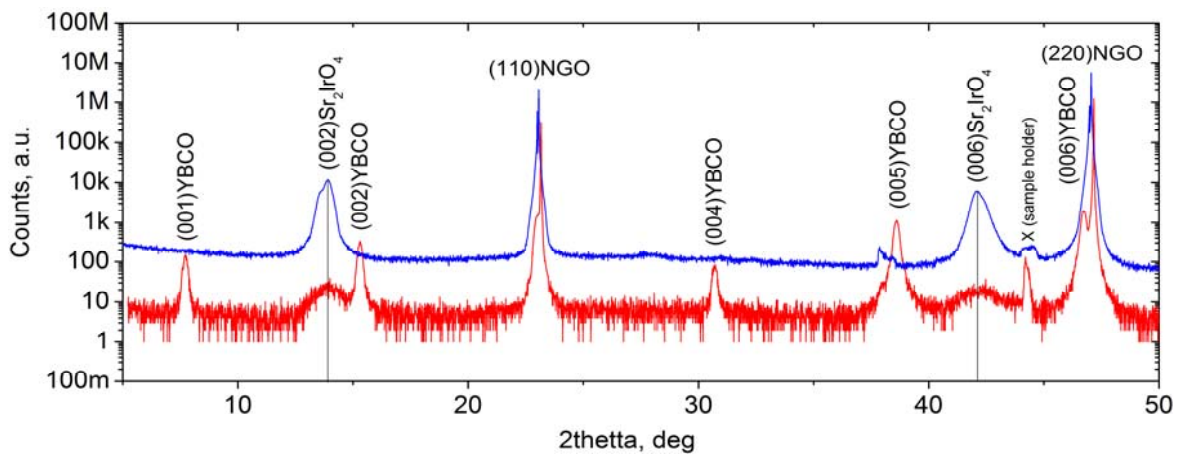

Fig.1. X-ray diffraction patterns of Bragg reflections for 20 nm thick Sr$_2$IrO$_4$ film (blue), and of heterostructure Au/Sr$_2$IrO$_4$/YBa$_2$Cu$_3$O$_x$ (red) grown on NdGaO$_3$. In heterostructure, thicknesses were 10 nm, 7 nm, 60 nm for Au Sr$_2$IrO$_4$ , and YBa$_2$Cu$_3$O$_x$, correspondingly

A protective Au thin film was deposited *in situ* as well at 30° C in the PLD chamber. The proximity effect between Nb and Au ensured the penetration of the superconducting order parameter through the Au film up to the Au/SIO interface as in the case of Au/YBCO junction [41]. The crystalline parameters of the films were determined using the 4-circle X-ray diffractometer, measuring X-ray diffractograms of 2Θ/ω scan and rocking curves. The X-ray diffraction patterns of Bragg reflections of a reference SIO film and the heterostructure Au/SIO/YBCO film are shown on Fig. 1. The YBCO film was grown on a (110) NGO substrate epitaxially with the c-axis perpendicular to the substrate plane, which was confirmed by the recorded spectra [37]. The epitaxial SIO film grown on YBCO is also *c*-oriented. It was found from Fig. 1 that the crystallographic lattice parameter *c* of the observed film (for 17 nm thickness) is 1.283 nm, which is close to the published value for $Sr_2IrO_4$ single crystals ($c$ = 1.290 nm) [42].

The results of measurements of the dependence of the conductivity of SIO films on temperature indicate the activation character of conductivity, the characteristic of an insulator (see Supplemental Material). The superconducting Nb film was deposited by magnetron sputtering in an argon atmosphere at room temperature just before the process of mesa fabrication. The temperature dependences of the resistance of the heterostructures were measured before the deposition of Nb, and afterward once again with a current flowing in-plane, in parallel to the substrate. After deposition of the Nb film, the ion etching through a photomask was used in order to form the geometry of the bottom electrode. At this step of the mesa fabrication, a decrease of the YBCO superconducting transition temperature from 90 K to 80-85 K was registered, which could be associated with oxygen depletion.

Nb/Au/SIO/YBCO mesa-heterostructures (MS) with sizes from $A$=10x10 to 50x50 μm$^2$ (total 5 MS on one chip) were formed using optical lithography, reactive ion-plasma and ion-beam etching at a low ion accelerating voltages. The $SiO_2$ protective insulator layer was deposited by RF sputtering and makes the dc current to flow in the perpendicular direction to the MS layers. An additional Nb film with a thickness of 200 nm was sputtered providing superconducting current transport through the dc wiring. Contact pads were made of gold for 4-point $I$-$V$ curve measurements: two ones over the YBCO electrode and two others over the Nb electrode (see Fig. 2).

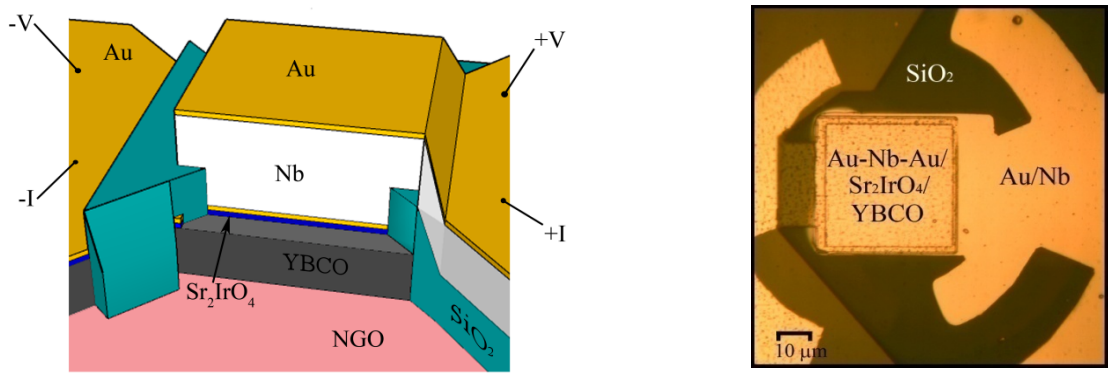

Fig.2. The schematic 3D view (on the left) and the top-view (on the right) of the mesa-heterostructure Nb/Au/SIO/YBCO. The insulating $SiO_2$ layer is used for the separation the top and bottom electrodes and realization electrical transport along the $c$-axis.

Fig. 3a shows the temperature dependence of the MS resistance $R(T)$ for the MS with in-plane size 40x40 μm$^2$ and SIO thickness $d$=7nm. Although the temperature dependences of the SIO film had an activated character of the resistivity (see Supplemental Material) at relatively high temperatures, above the superconducting transition temperature, the resistances of Nb and

YBCO electrodes gave the main contribution to R at $T>T_C$. After finish Nb wiring pattering the transition temperature of the YBCO electrode in the MS has reduced again to $T_C \approx 61$ K (Fig.3a), which could be explained by the influence of ion-beam etching. Some reduction of $T_C$=8.4 K for Nb film also took place. There are three interfaces in MS: Nb/Au, Au/SIO and SIO/YBCO. The Au interlayer in between of Nb and YBCO layers prevents the oxidation of Nb and considerably weaken the oxygen depletion of the SIO/YBCO bilayer. Our investigation of the Nb/YBCO interface (without Au layer between superconductors) reveals much higher values of $R \cdot A$ =0.1–1 $\Omega \times cm^2$ (A is the in-plane area of MS), and no superconducting current has been observed in Nb/YBCO structures [43].

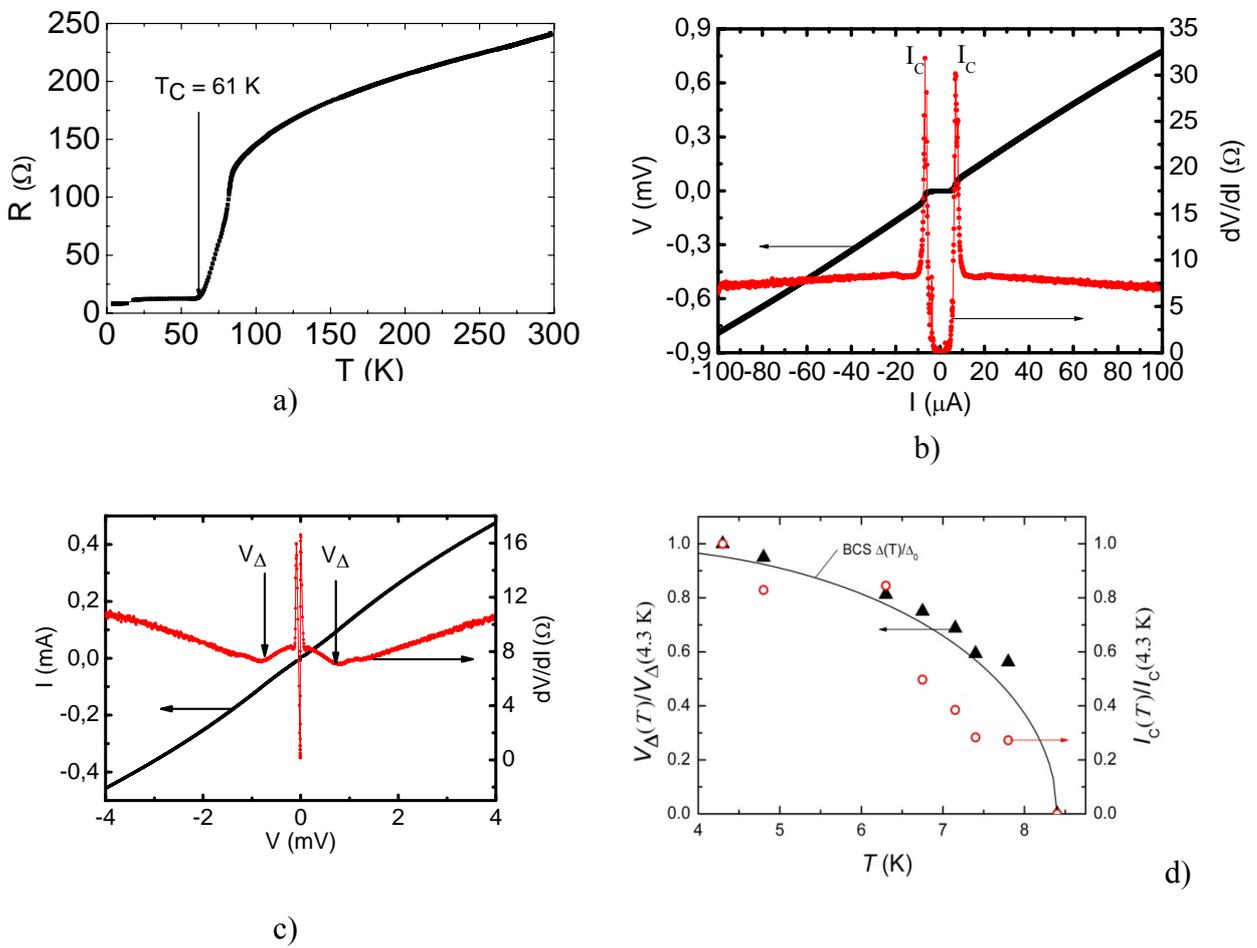

Fig.3. a) Temperature dependence of the resistance of MS Nb/Au/SIO/YBCO with size 40x40 μm² and SIO film thickness d=7nm b) I-V curve and the dependence of differential resistance dV/dI versus I, allowing to estimate the value of critical current $I_C$ at T=4.2K, c) I-V curve and dV/dI versus V, which shows singularities caused by the energy gap of the Nb electrode at T=4.2K, d) temperature dependences of the normalized critical current $I_C$ and gap voltage $V_\Delta$, where $V_\Delta$(4.3 K) = 0.8 mV. A solid line is the ordinary BCS dependence of the energy gap versus temperature.

The normal resistance of the Nb/Au interface $R_{Nb/Au}$ was in μΩ range and $R_{Nb/Au}A \sim 10^{-11} \Omega \times cm^2$, which corresponds to the Nb/Au interface transparency $\mathfrak{I}_{Nb/Au} \approx 1$ [41]. The averaged value of the normal resistance for 4 MS on the one chip with SIO thickness d=7 nm was $R_N A \approx 100$ $\mu\Omega \times cm^2$ at $T$=4.2 K. Note, the contribution of resistivity from SIO interlayer, estimated from measurements of SIO film $\rho \times d = 7 \times 10^3 \mu\Omega \times cm^2$ ($d$=7 nm) should be much larger (see Supplemental Material). Thus, it allows us to argue that the tunneling through the SIO film in the MS with a low barrier transparency $\mathfrak{T}= 3 \cdot 10^{-5}$ is the main mechanism for electrical transport.

Figures 3b and 3c show $I$-$V$ curves at $T$=4.2 K of the same MS, plotted in two different ranges of dc bias. Fig.3b shows the $I$-$V$ curve and the differential resistance $R_D$=d$V$/d$I$ from which we evaluated the values of the critical current $I_C$, and Fig.3c shows how we did evaluate the gap voltage $V_\Delta$ caused by the energy gap of the Nb electrode from a singularity on the $I$-$V$ curve. The temperature dependences of $I_C$ ($T$) and $V_\Delta$($T$) are given in Fig.3d. At temperatures near Nb film critical temperature $T_{CNb}$ the critical current $I_C$ is small and it was determined from $dV/dI$($I$) functions since the influence of fluctuations resulted in "rounded" $I$-$V$ curves. In this case, an approach described in [44] (see also Supplemental Material), which takes into account the external low-frequency fluctuations [45] was used. In Table 1 the dc parameters of 4 MS on the same chip with thickness $d$=7 nm are presented. Three samples with $L$=50, 40 and 30 μm show relatively good reproducibility of the $I_C R_N$ product and $R_N A$. The smaller one, $L$=20 μm, had $R_N A$ which differs not too much, but had 3 times smaller $j_C$. The 5[th] sample with $L$=10 μm had much higher $R_N A$ and it was omitted.

Table 1

| $L$ (μm) | $I_C$ (μA) | $R_N$ (Ω) | $R_N A$(Ω cm$^2$) | $j_C$ (A/cm$^2$) | $I_C R_N$ (μV) | $\lambda_J$ (μm) |
|---|---|---|---|---|---|---|
| 50 | 6.5 | 5.0 | 125 | 0.26 | 32 | 725 |
| 40 | 6.0 | 7.1 | 114 | 0.38 | 43 | 600 |
| 30 | 3.0 | 10.4 | 94 | 0.33 | 31 | 645 |
| 20 | 0.5 | 20.7 | 83 | 0.12 | 10 | 1065 |

$L$ is the planar size of the MS, $A=L^2$, $I_C$ is critical current, $R_N$ is normal resistance measured at $V$=0.8 mV, $j_C= I_C/A$ is critical current density, the characteristic resistance of MS $R_N A$, $\lambda_J = (\hbar/2e\mu_0 d j_C)^{1/2}$ is Josephson penetration depth at $T$=4.2 K, external magnetic field $H$=0.

In MS the s-wave Nb/Au superconducting electrode contacts via the SIO barrier with the YBCO superconductor which order parameter is usually described as a superposition of d-wave ($\Delta_d$) and s-wave ($\Delta_s$) components: $\Delta(\theta)=\Delta_d\cos2\theta+\Delta_s$, where θ is the angle between the quasiparticle momentum and the $a$-axis of the YBCO [41]. In this case, the superconducting current-phase relation (CPR) may differ from the sinusoidal one. Particularly, it happens for the hybrid s- and d-wave (S/D) superconducting junctions for the transport along the *c*-direction $D_{001}$ [46, 47]. The CPR in S/$D_{001}$ junction could be represented as:

$$I_s(\varphi)=I_{c1}\sin\varphi+I_{c2}\sin2\varphi \qquad (1)$$

where $I_{c1}$ and $I_{c2}$ are amplitudes of the critical current for the first and the second harmonics, and the ratio $q = I_{c2}/I_{c1}$ is used as a characteristic parameter for the second harmonic in the CPR. Note, in presence of the second harmonic in the CPR the measured $I_C$ does not coincide with the $I_{c1}$, and the relationship of $I_C/I_{c1}$ and $q$ is given in [41]. If $q \leq 0.5$ the difference between $I_{c1}$ and $I_C$ is less than 20%. The first harmonic $I_{c1}$ originates from the minor s-wave component of the superconducting order parameter in YBCO ($\Delta_s$). It has been shown theoretically [48] that in S/$D_{001}$ junctions the first harmonic of CPR in the case $\Delta_d \gg \Delta_s$, $\Delta_{Nb}$ ($\Delta_{Nb}$ is Nb superconducting gap) looks as follows:

$$I_{c1}R_N \approx \Delta_s\Delta_{Nb}/(e\Delta_D^*) \qquad (2)$$

where $\Delta_D^* = \pi\Delta_d[2\ln(3.56\Delta_d/k_BT_{cNb})]^{-1}$.

Fig. 4 shows the family of *I-V* curves at the fixed magnetic field *H* and dependence of the critical current on the external magnetic field for the MS with $L$=50 μm (see other parameters in Table 1). The experimental $I_C(H)$ dependence and the theoretical Fraunhofer pattern $I_C(H)=I_0|\sin(\pi H)/\pi H|$, [49] are shown in Fig.4b. The expected magnetic field which corresponds to the first minimum is $H_1 = \Phi_0/\mu_0 d_J L \simeq 4$ Oe, where $d_J=d+\lambda_{Nb}\tanh(d_{Nb}/2\lambda_{Nb})+\lambda_{YBCO}\tanh(d_{YBCO}/2\lambda_{YBCO})$, $d$=7 nm is thickness of SIO barrier, $L$=50 μm is junction size and $\lambda_{YBCO}$=150 nm, $\lambda_{Nb}$= 90 nm are London penetration depths for YBCO and Nb at 4.2 K, correspondingly. From Fig.4b it is seen that experimental $H_1$ is smaller than predicted one, and the half-width at the half-height of the central peak is narrower about 1.5 times than calculated for the case of a tunnel junction [50] with the geometry and magnetic field direction as in our experiment. An enhanced sensitivity of the superconducting current to magnetic field has been observed earlier for Josephson junctions with the antiferromagnetic $Ca_{0.5}Sr_{0.5}CuO_2$ interlayer [51] and was described by a specific orientation of a thick antiferromagnetic interlayer placed in between of the s-wave superconducting electrodes [52] which leads to the long-range spin-singlet Josephson effect.

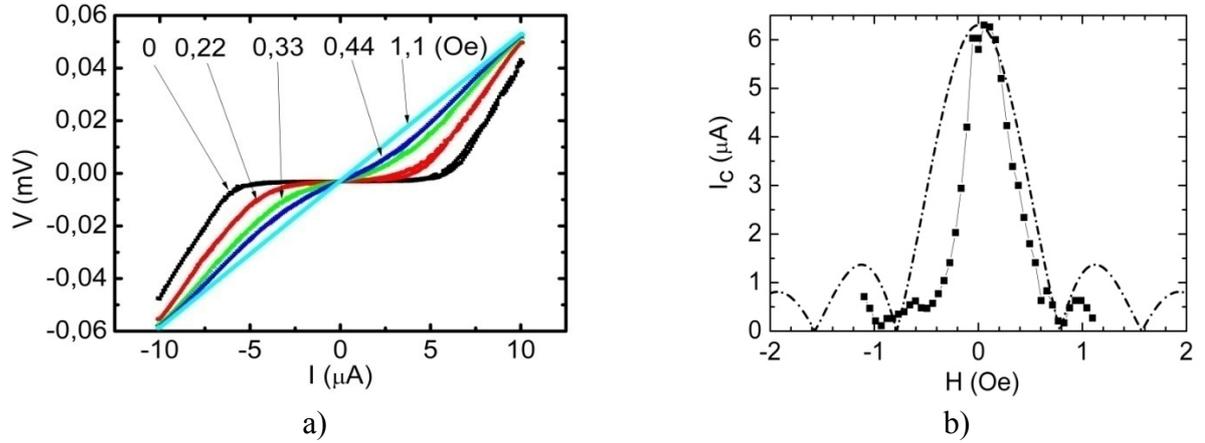

Fig.4 a) Family of *I-V* curves at *T*=4.2K and different magnetic field in range 0-1.1 Oe for MS with *L*=50μm and *d*=7 nm. b) The critical current versus the magnetic field (black squares). The calculated Fraunhofer pattern (shown by dash dot line) $I_C(H)=I_0|\sin(\pi H)/\pi H|$, where $H=\Phi/\mu_0 d_J$ for Josephson junction with *L*=50 μm and *d*=7 nm.

The *d*-wave component of the YBCO superconducting order parameter ($\Delta_d$) promotes the unconventional superconducting CPR of the junctions (see Eq. (1)) [41] and gives the 2$^{nd}$ harmonic

$$I_{c2}R_N \approx \mathcal{T}\Delta_{Nb}/e, \qquad (3)$$

where $\mathcal{T}$ is the barrier transparency.

The ac Josephson effects in MS was revealed by measurements of *I-V* curves under electromagnetic radiation in mm frequency range. Since the characteristic frequency for MS $f_C=(I_C R_N)2e/h$ is in GHz-range the investigation was carried out both at $f_e$=38 GHz and at $f_e$=50 GHz. Fig.5a shows an example of the voltage dependence of d*V*/d*I* of MS at $f_e$=50 GHz, demonstrating the existence of integer and fractional Shapiro steps arising due to synchronization between Josephson self-oscillations in MS and the external microwaves at voltages $V_{n,m}=(n/m)hf_e/2e$. Minima of d*V*/d*I*(*V*) point on voltage positions $V_{n,m}$ of Shapiro steps. Fig.5b show dependences of normalized voltage $2eV_n/hf_e$ versus integer number n=m=1 of Shapiro step obtained at two frequencies $f_e$=38 GHz and $f_e$=50 GHz of microwave irradiation. Excellent correspondence to the Josephson voltage-frequency relation is seen. As seen from Fig.5a the fractional Shapiro steps exist and may indicate the presence of the second harmonic in CPR *q*≠0 [see Eq.(1)].

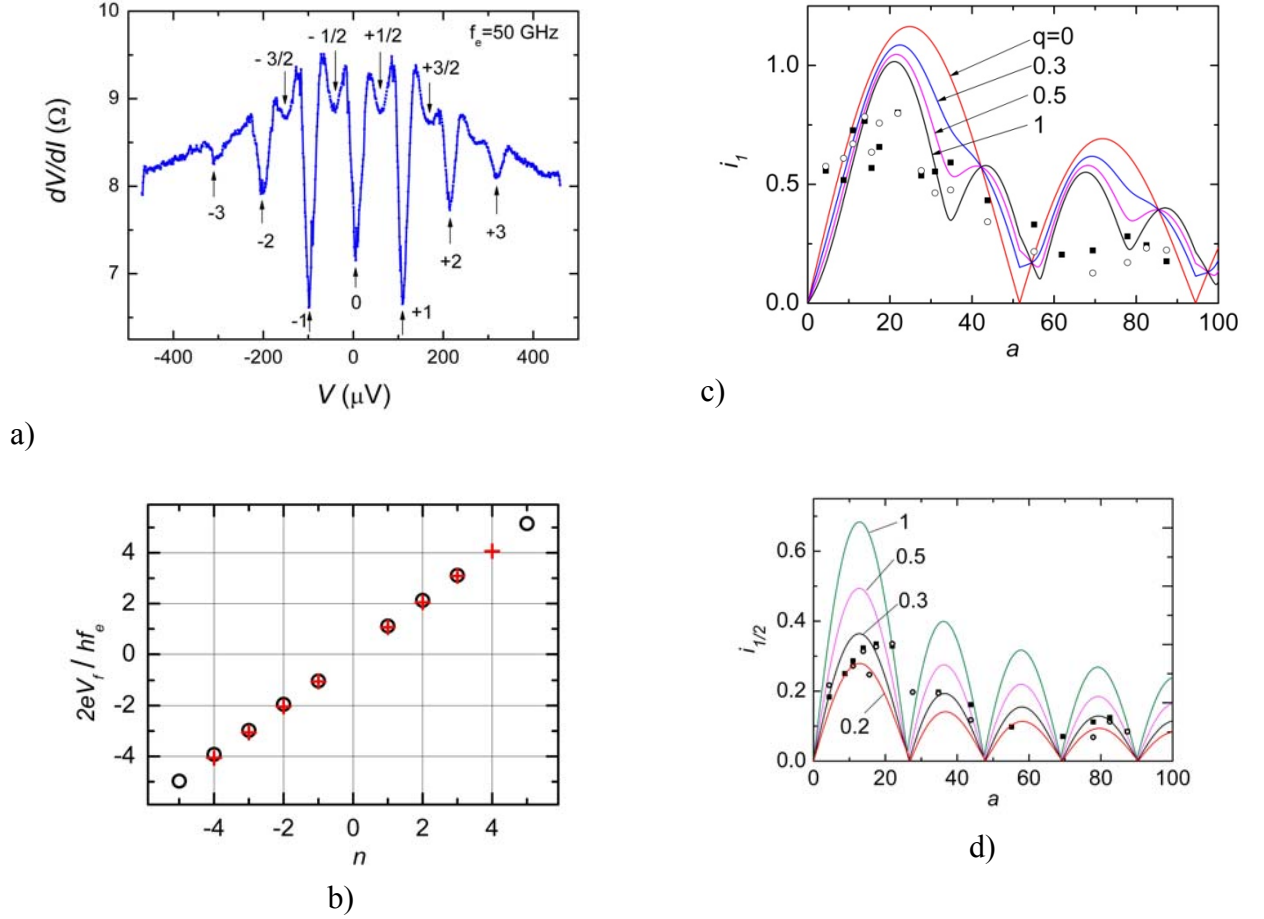

Fig.5 a) Voltage dependence of the differential resistance d$V$/d$I$ for the MS with $d$=7 nm, $L$=40 μm, $T$=4.2 K, at microwave radiation 50 GHz. The arrows and numbers indicate the position of integer and fractional Shapiro steps. The number "0" corresponds to the critical current. The microwave power was adjusted for the demonstration of fractional Shapiro steps. b) Dependences of the position on voltage axis for integer Shapiro step $V_n$ =n2e$V$/h$f_e$ versus Shapiro step integer number $n$ for two frequencies $f_e$=38 GHz (open circles) and $f_e$=50 GHz (crosses). c) The normalized amplitude of Shapiro steps $i_1$=$I_1(a)/I_C(0)$ versus normalized microwave current. $a$=$I_{MW}/I_C(0)$. Open and filled symbols belong to positive and negative dc voltage biasing, correspondingly. Theoretical curves were calculated taking $a$=$I_{MW}/I_C$ as a fitting parameter for a ratio $q$= $I_{c2}/I_{c1}$=0, 0.3, 0.5, 1 and McCumber parameter $\beta_C$=1. d) Normalized amplitudes of half-integer Shapiro steps $i_{1/2}$=$I_{1/2}(a)/I_C(0)$. Theoretical curves were calculated for $q$= 0.2, 0.3, 0.5, 1 and $\beta_C$=1.

The information on the CPR in MS could be obtained from the dynamics of Shapiro steps varying the power of microwave irradiation in conditions of high frequency limit $f_e$>$f_C$ [51]. Dependences of the first $i_1$=$I_1/I_C$ ($n$=$m$=1), and the fractional (half-integer) $i_{1/2}$=$I_{1/2}/I_C$ (n=1, m=2) Shapiro steps on the normalized microwave current $a$=$I_{MW}/I_C(0)$ at $f_e$=50 GHz are shown in Fig.5c and Fig.5d, correspondingly. Calculated dependences of $i_1(a)$ and $i_{1/2}(a)$ for different values of $q$ and fixed McCumber parameter $\beta_C$=$f_C/f_{RC}$=1 ($f_{RC}$=1/$\pi R_N C$) using the modified RSJ model [51, 53] are also presented in Fig.5c and Fig.5d. However, dependence $i_1$=$I_1(a)/I_C(0)$ in Fig.5c hardly helps to estimate $q$-factor, at the same time as seen from Fig.5d the maximum of half-integer Shapiro step has max($i_{1/2}$)≈0.3 and fits well enough to the case $q$=0.3.

Estimated contribution of the second harmonic caused from the d-wave symmetry for the case of S/D$_{001}$ junction with the same electrical parameters as the discussed MS with $L$=40 μm gives $I_{c2}$≈3.4 nA for $\mathfrak{J}$ =3·10$^{-5}$, $\Delta_{Nb}/e$=0.8 mV and $R_N$ =7.1 Ω at $T$=4.2 K (see Eq. (3)). Using the theoretical dependence for ratio $I_{c1}/I_C$ versus $q$ [41] and the estimated "d-wave" contribution of $I_{c2}$, it gives negligibly small $q \approx 6 \cdot 10^{-4}$. The existence of Josephson tunneling current through the relatively thick insulating barrier $d$=7 nm with low transparency could not be explained by ordinary Josephson effect as in S-I-S tunnel junctions. At the same time, the deviation of the CPR from the sinusoidal type may originate from an appearance at the interface SIO/YBCO energy bands related to the coherent Andreev reflections giving rise of superconducting current through antiferromagnetic insulator with SOI [39, 40, 54]. The theoretical simulation [40] shows that the interface of cuprate superconductor with an Ir oxide layer of Sr$_2$IrO$_4$ could exhibit both helical Majorana fermions and zero-energy flat edge states. Indeed, the MS demonstrated the zero-bias conductivity peak (ZBCP) at both $T$=4.2 K and higher temperatures $T>T_{CNb}$ which could be associated with Andreev bound states at the interface (see Supplemental Material). However, the origin of the ZBCP in the MS requires additional studies.

In conclusion, Nb/Au/Sr$_2$IrO$_4$/YBa$_2$Cu$_3$O$_x$ mesa-heterostructures have been fabricated with epitaxial bilayer consisting of films Sr$_2$IrO$_4$ and YBa$_2$Cu$_3$O$_x$ and superconducting current has been observed. The critical current of the mesa-heterostructure increased with decreasing temperature is proportional to the voltage of the gap singularity of Nb film. The $I_C(H)$ dependence showed a sharp central peak and minor oscillating behavior deviated from usual for ordinary Fraunhofer pattern for Josephson junctions. Under the influence of electromagnetic radiation of the millimeter-frequency range, the Shapiro steps were observed at voltages both at multiple and fractional quantities (n/m)h$f_e$/2e, indicating the deviation of the current-phase relation from the sinusoidal type and the presence of the second harmonic.

The authors are gratefully acknowledge I.V. Borisenko, L.V. Filippenko, and D. Winkler for the help and useful discussions, GAO and AVS would like to acknowledge COST Action MP1308 – TO-BE for Short-Term Scientific Mission (STSM) program.

Supplemental Material for

"Observation of superconducting current in a mesa-heterostructure with an interlayer of the strontium iridate film with strong spin-orbit interaction"

A.M. Petrzhik[1], K.Y. Constantinin[1], G.A. Ovsyannikov[1], A.S. Grishin[1], A.V. Shadrin[1,2], Yu.V. Kislinski[1], G. Christiani[3], G.Logvenov[3]

[1]Kotel'nikov IRE RAS, Mokhovaya 11-7 Moscow, Russia.

[2] Moscow Institute of Physics and Technology, Dolgoprudny, Moscow Region, 141701, Russia.

[3]Max Planck Institute for Solid State Research, Stuttgart, 70569, Germany.

*The resistivity of $Sr_2IrO_4$ film*

We studied resistivity of $Sr_2IrO_4$ (SIO) film deposited on the following substrates: $(100)SrTiO_3$ (STO), $(110)NdGaO_3$ (NGO), $(100)La_{0.3}Sr_{0.7}Al_{0.65}Ta_{0.35}O_3$ (LSAT), and $(100)LaAlO_3$ (LAO), keeping the same SIO thickness $d=34$ nm. Temperature dependences of resistivity were measured using the Keithley 6517B electrometer and the Neocera LTC temperature controller. Fig.1S shows temperature dependences of resistivity $\rho(T)$ and $\rho(1000/T)$ for SIO films on LSAT, STO, and LAO substrates. A "linear" part of $\rho \sim 1/T$ is associated with activation Arrhenius type of resistivity $\rho=\rho_0 \exp(\Delta E_g/2k_B T)$, where dielectric gap $\Delta E_g =200–230$ meV [1S], $k_B$ is the Boltzmann's constant. The values of room-temperature resistivity $\rho_0$ of as-grown films were by several orders greater than the values reported in [2S] which could be related to the difference in the regimes of SIO film deposition. As seen from this figure, the activation mechanism characterized well the $\rho(T)$ dependence at intermediate temperatures. Further decreasing the temperature (above the dotted line in Fig.1S) the resistivity exhibited significant rise and the total resistance of the film became $R >50$ GΩ (the upper limit of our measurements). Thus, at $T<100$K (roughly) an additional mechanism is switched on which could be described by a simple 3D variable-range hopping (VRH). Temperatures where this mechanism becomes a dominant correspond to a linear dependence of $\rho$ vs. $T^{-1/4}$. The hopping range was estimated as $b = a(T_0/T)^{1/4}$, where $a$ is the localization radius, and amounted $b =13–17$ nm which did not exceed the thicknesses of SIO interlayer in the MS.

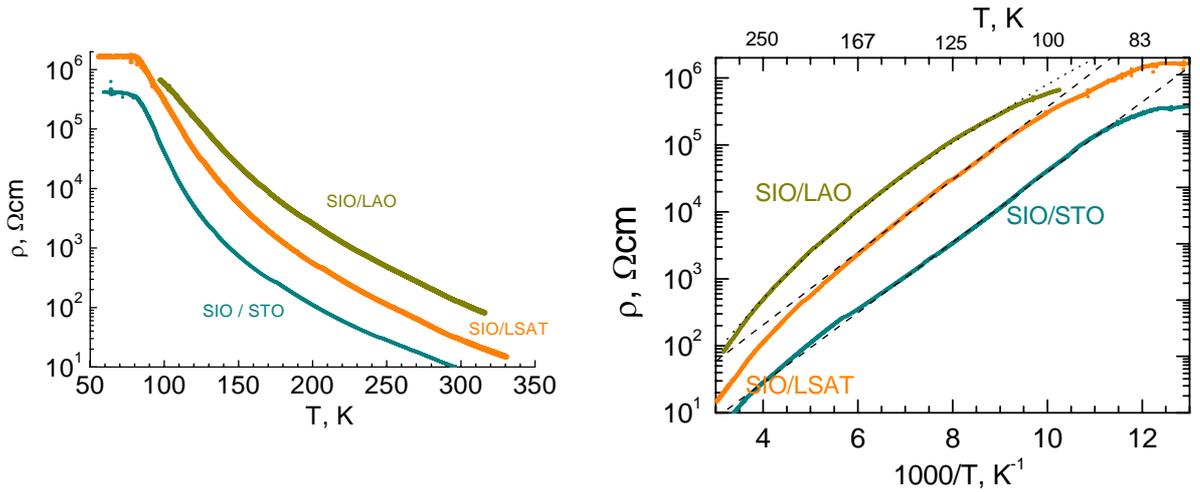

Fig.1S. Temperature dependence of resistivity for $Sr_2IrO_4$ films deposited on LAO, LSAT and STO substrates; left panel shows $\rho(T)$ in semi-log scale, right panel shows $\rho(1000/T)$. Activation $\rho=\rho_0\exp(\Delta E_g/2k_BT)$ dependences are shown by the dotted lines.

*Critical current and Shapiro steps determination*

In order to determine the values of critical current $I_C$ of MS under influence of the fluctuations, we analyzed the dependences of differential resistances $dV/dI$ versus biasing current $I$. Fig. 2S shows an example of a family of $dV/dI(I)$ at magnetic fields $H$=0.44, 0.55, and 0.82 Oe. Influence of fluctuations results in "rounded" I-V curves without sharp voltage drop. In this case the parameter of a low frequency noise, $I_f$, was introduces [3S - 5S] for a Josephson junction biased by the current source and fluctuations at frequencies below $f_C=(2e/h)I_CR_N$. If $I_C > I_f$ then critical current $I_C$ is easy to determine from the interval between sharp maxima of $dV/dI$ on the current $I$-axis. In the case $I_C<I_f$ as shown in Fig.2S the $dV/dI$ peaks are rounded and the approach [3S- 5S] was used which takes into account the low frequency noise current, $I_f$. From a comparison of the experimental and theoretical dependences of $dV/dI$ vs. $I$ the effective $I_f$ was determined for a family of I-V curves measured at the same conditions (at the same level of $I_f$) [3S-4S]. Fig.3S shows theoretical functions $r$ vs. $2I_C/I_f$ ratio for: $r=1+R_{Dmax}/R_{Dmin}$, $r=R_{Dmax}/R_N$ and $r=R_{Dmin}/R_N$, where $R_N$ is the normal state resistance of Josephson junction, $R_{Dmin}= dV/dI$ at $I$=0, $R_{Dmax}$ is determined as local maximum of $dV/dI(I)$ (e.g., see Fig.2S). In the case when $R_N$ could not be determined with the required accuracy (say, $R_N$ is varied under microwave irradiation or magnetic field) function for $r=1-R_{Dmax}/R_{Dmin}$, excluding $R_N$, was used. An example, where $R_N$ is hard to determine as an asymmetry in conductance $G(V)$ exists is shown in Fig.4S which shows also the zero bias conductance anomaly. Note, the I-V curve in the vicinity of the Shapiro step has the same hyperbolic shape as in the stationary case [4S, 5S]. So, this approach is also applicable to determine the Shapiro step amplitudes $I_{n/m}$.

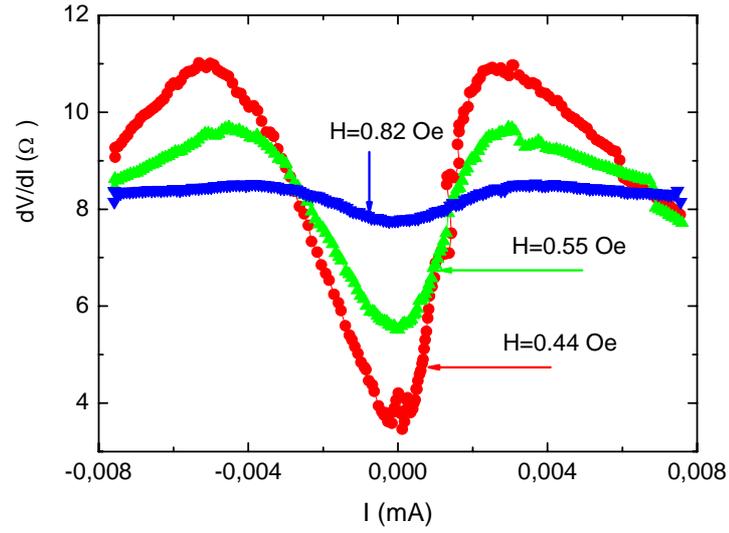

Fig.2S. A family of d$V$/d$I$ vs. $I$ for magnetic field levels $H$=0.44, 0.55, and 0.82 Oe for MS with $d$=7nm and sizes 50x50 μm$^2$

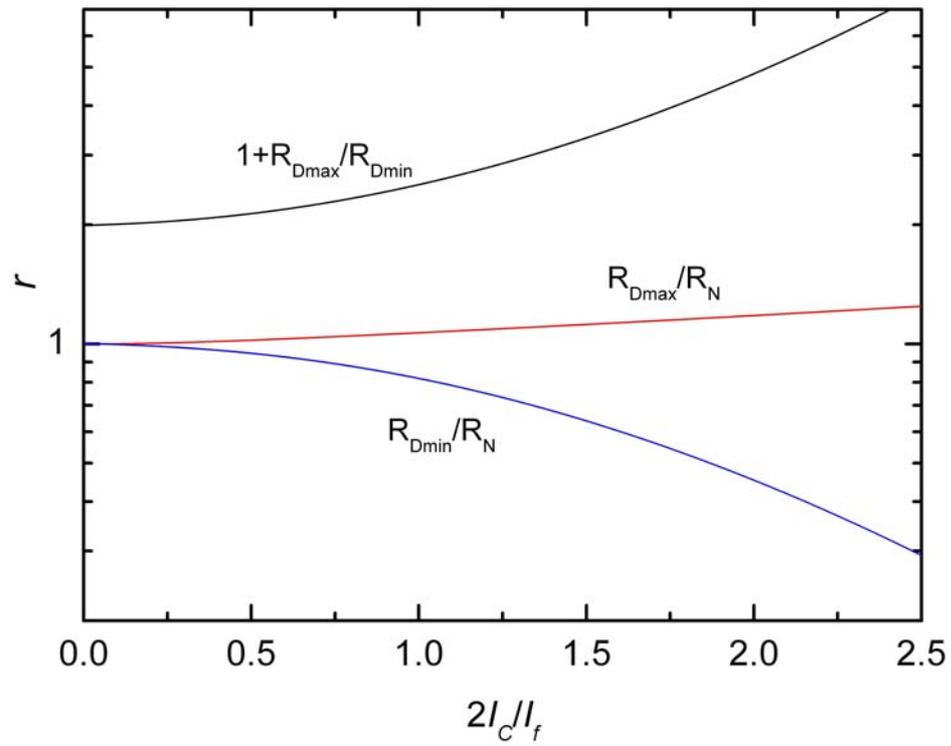

Fig.3S. Theoretical functions for $r$ vs. $2I_C/I_f$: from top to bottom $r$=1+$R_{Dmax}/R_{Dmin}$, $R_{Dmax}/R_N$, $R_{Dmin}/R_N$.

*Zero bias conductance anomaly.*

The zero bias conductance anomaly is seen on the voltage dependence of conductivity *G(V)* at the both temperatures, *T*=4.2 K when MS is superconducting, and at *T*=15.3 K when only YBCO electrode is in superconducting state (Fig.4S). Asymmetry of *G(V)* seen also at relatively low voltages *V*<10 mV which hardly could be explained by an impact of a Schottky barrier. This may be caused by an asymmetry of a spin-polarized charge transport through the SIO barrier, but requires additional studies.

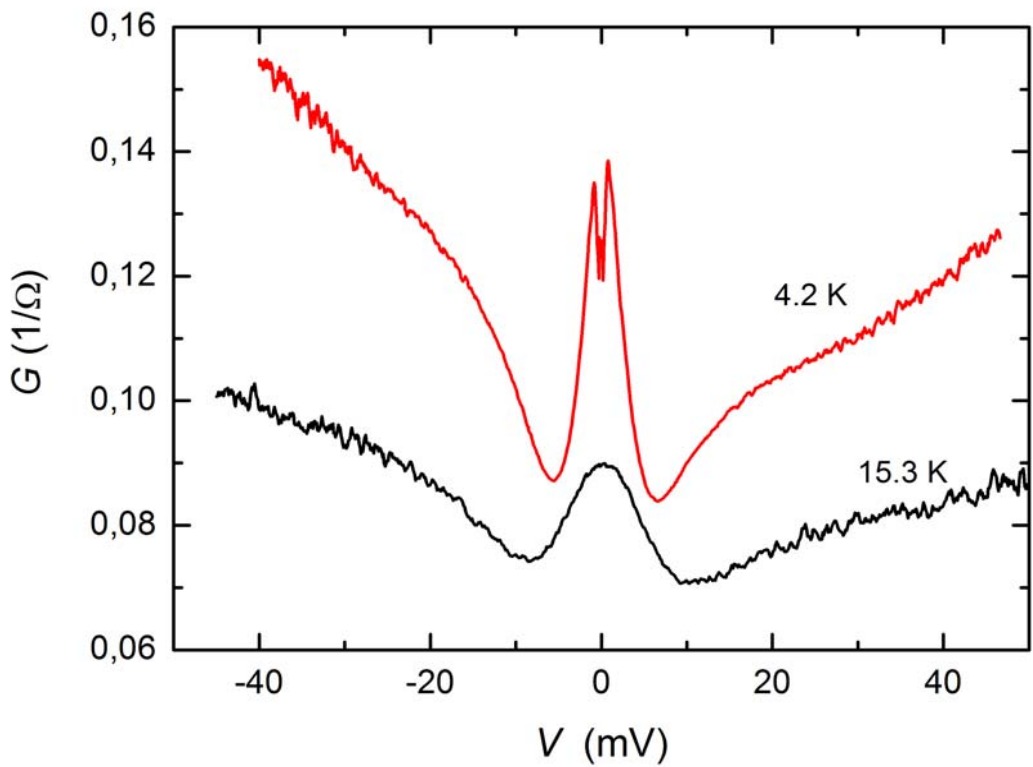

Fig.4S. Voltage dependence of conductance *G(V)* for MS with *d*=7 nm, 40x40 μm$^2$ at *T*=4.2 K and *T*=15.3 K.